\documentclass[12pt]{iopart}

\usepackage{graphicx}
\usepackage{subfigure}
\begin{document}

\begin{flushleft}
\Large{\textbf{Ordering of localized electronic states in multiferroic TbMnO$_3$: a soft X-ray resonant scattering study}}
\end{flushleft}
\author{T R Forrest$^1$, S R Bland$^2$, S B Wilkins$^3$, H C Walker$^1$, T A W Beale$^2$,
P D Hatton$^2$, D Prabhakaran$^4$, A T Boothroyd$^4$, D Mannix$^5$, F Yakhou$^6$ and D F McMorrow$^1$}

\address{$^1$London Centre for Nanotechnology, University College London, Gower Street, London, WC1E~6BT, United Kingdom}
\address{$^2$Department of Physics, University of Durham, Rochester Building, South Road, Durham, DH1~3LE, United Kingdom}
\address{$^3$Brookhaven National Laboratory, Condensed Matter Physics and Material Science Department,
Bldg \#501B, Upton, NY 11973-5000, United States of America}
\address{$^4$Department of Physics, University of Oxford, Clarendon Laboratory,
Parks Road, Oxford, OX1~3PU, United Kingdom}
\address{$^5$Institut N'{e}el, CNRS-UJF, BP166, 38042 Grenoble, France}
\address{$^6$European Synchrotron Radiation Facility, BP220, 38043 Grenoble, France}
\ead{t.forrest@ucl.ac.uk}
\begin{abstract}
Soft X-ray resonant scattering (XRS) has been used to observe
directly, for the first time, the ordering of localized electronic states
on both the Mn and Tb sites in multiferroic TbMnO$_3$. Large
resonant enhancement of the X-ray scattering cross-section were
observed when the incident photon energy was tuned to either the Mn
$L$ or Tb $M$ edges which provide information on the Mn $3d$ and Tb
$4f$ electronic states, respectively. The temperature dependence of the XRS
signal establishes, in a model independent way, that in the
high-temperature phase (28~K $\leq$ T $\leq$ 42~K) the Mn $3d$
sublattices displays long-range order. The Tb $4f$ sublattices are
found to order only on entering the combined ferroelectric/magnetic
state below 28~K. Our results are discussed with respect to recent
hard XRS experiments (sensitive to spatially extended orbitals) and
neutron scattering
\end{abstract}

\section{Introduction}
Magnetoelectric multiferroics are materials that simultaneously display ferroelectric and magnetic 
long-range order \cite{eerenstein}. Consequently, they
are of considerable interest both from a fundamental point of view,
and for the potential that they offer in the field of spintronics
\cite{Fiebig,Spaldin}. Of particular importance has been the recent
discovery of multiferroic behaviour in a diverse range of compounds where
the multiferroic state takes the form of the coexistence of
ferroelectricity and antiferromagnetism, often with a large coupling
between the two \cite{Hill,Cheong}. Indeed the burgeoning  interest
in multiferroics can be traced to the pioneering work by Kimura \textit{et
al.} who demonstrated a giant magnetoelectric effect in TbMnO$_3$
where the electric polarization may be switched by applying a
magnetic field \cite{KimuraTMO}. More recently it has been shown
how an applied electric field can be used to manipulate the magnetic
domain distribution \cite{yamasaki}.

For TbMnO$_3$ (space group \textit{Pbmn}) the Mn$^{3+}$ magnetic moments
first order below $T_{N1}$=42~K. It has been reported, based on recent neutron diffraction
data, that below this temperature the moments are polarized along the {\bf b} direction
with a modulation wavevector (0 q$_{Mn}$ 0), q$_{Mn}$ $\approx$ 0.29 b*. Below $T_{N2}$=28~K the magnetic structure adopted by the Mn
sublattice becomes non-collinear, forming a cycloid in the {\bf
b}-{\bf c} plane \cite{kenzelmann} and, at exactly the same temperature,
a ferroelectric polarization along the {\bf c} direction is observed. 
By comparing the magnetic structures above and below
$T_{N2}$, an elegant and appealing model was proposed whereby the
ferroelectric transition is driven by a loss of inversion symmetry
at the Mn sites as the magnetic structure changes from collinear to
non-collinear. The same study also proposed that the Tb magnetic moments
are disordered in the collinear phase, and become polarized along the {\bf a} direction 
on cooling into the cycloidal phase.

X-ray resonant scattering (XRS) has much to offer the study of
multiferroics in general \cite{ewings,Prokhnenko,yang,koo}, and
TbMnO$_3$ in particular. It is an element and electron shell
specific technique, which in the case of TbMnO$_3$ makes it possible
to study any ordering of the Mn and Tb sublattices separately.
Neutron diffraction by contrast measures the scattering from the
sum of the separate contributions. Moreover, XRS is capable of
providing information on the presence of any multipolar order
\cite{matteo}. This includes multipoles with time-odd, parity-odd
symmetry that may characterize the combined magnetic and
ferroelectric state displayed by TbMnO$_3$ and other related
multiferroics \cite{VanAken}. Recently the results of a number of XRS studies of
TbMnO$_3$ have been reported, all performed in the hard part of the
X-ray spectrum above 3 keV \cite{mannix,argyriou}. The edges
accessed in this part of the spectrum are the Mn $K$ and Tb $L_3$
edges which, for the dominant electric dipole resonances observed, provide
information on the ordering of the $4p$ and $5d$ extended band
states at the Mn and Tb sites, respectively. Here we highlight two
of the main results of this X-ray work \cite{mannix}. The first is the surprising
observation of a large polarization of the Tb $5d$ states in the
collinear phase, where according to modeling of the neutron data
the Tb magnetic moments are disordered. The second is the tentative
report that the $5d$ states develop an anapolar moment, {\textit i.e.} one with time-odd, parity-odd symmetry, in the
cycloidal phase.

In order to shed further light on the ordering of the different
electronic states in TbMnO$_3$ we have utilized XRS in the vicinity
of the Mn $L_2$ (649.9 eV: $2p_{3/2}\rightarrow 3d$) and $L_3$ (638.7 eV:
$2p_{5/2}\rightarrow 3d$) and the Tb $M_4$ (1276.9 eV:
$3d_{3/2}\rightarrow4f$) and $M_5$ (1241.1 eV:
$3d_{5/2}\rightarrow4f$) edges. The clear benefit of utilizing these
edges is that they provide information on the localized $3d$ and
$4f$ states. In this sense our soft X-ray study of TbMnO$_3$ is
complementary to both neutron diffraction and hard XRS studies.
There are, however, limitations as to what can be achieved with soft
X-rays. For example, when compared to the studies performed with photon energies above 3 keV, the Ewald sphere of reciprocal space is severely limited.

One complication encountered in any study of TbMnO$_3$, in addition
to the ordering of both the Tb and Mn sublattices, is the existence
of a complex magnetic domain structure. Four possible domain states
have been identified, which can be classified according to the
Miller indices ($h,k,l$) of the associated satellite peaks: $A$ ($h$+$k$=even, $l$=odd), $C$ ($h$+$k$=odd, $l$=even), $F$
($h$+$k$=even, $l$=even), and $G$ ($h$+$k$=odd, $l$=odd).
The relative population of domains is found to differ from study to
study \cite{Quezel,Blasco,Kajimoto,kenzelmann}, although the $A$
domain dominates in most cases reported. In our soft XRS study, the
limited range of reciprocal space available had the
consequence that only $F$-type satellite reflections fall within the
Ewald sphere at the Mn $L$ edges, while at the Tb $M$ edges $A$, $C$
and $F$ satellites can in principle be accessed.

\section{Experimental Details}
Single crystals of TbMnO$_3$ with dimensions $2\times2\times1$
mm$^3$ were grown at the University of Oxford using the flux growth
method. They were cut with either the [0~1~0] or [0~0.28~1] directions as the
surface normal and polished with 0.1$\mu$m diamond followed by 0.02$\mu$ Al$_2$O$_3$ pastes, to a
flat, shiny surface.
Experiments were carried out on both beamline 5U1 at the SRS, Daresbury Laboratory and ID08 at the European Synchrotron Radiation Facility. The former beamline was used to collect the temperature dependence of the scattering, while the latter was used for its superior flux and high incident photon energy resolution to determine the energy dependence of the scattering. These measurements were conducted in a similar fashion to that of Wilkins \textit{et al.} \cite{Wilkins03_1,Wilkins03_2,wilkinsLSMO} on both beamlines. The samples were mounted on the diffractometer with the surface normal and [001] direction lying within the scattering plane. In both cases, the diffraction plane was vertical. At 5U1 and ID08 the base temperatures achievable were 22~K and 19~K respectively. Due to experimental apparatus limitations it was not possible to measure the polarization of the scattered X-rays.

\section{Results and Discussion}
\begin{figure}
\centering \subfigure[]{\label{L$_3$ peak}
    \includegraphics[width=.25\textwidth,bb=10 170 570 600]{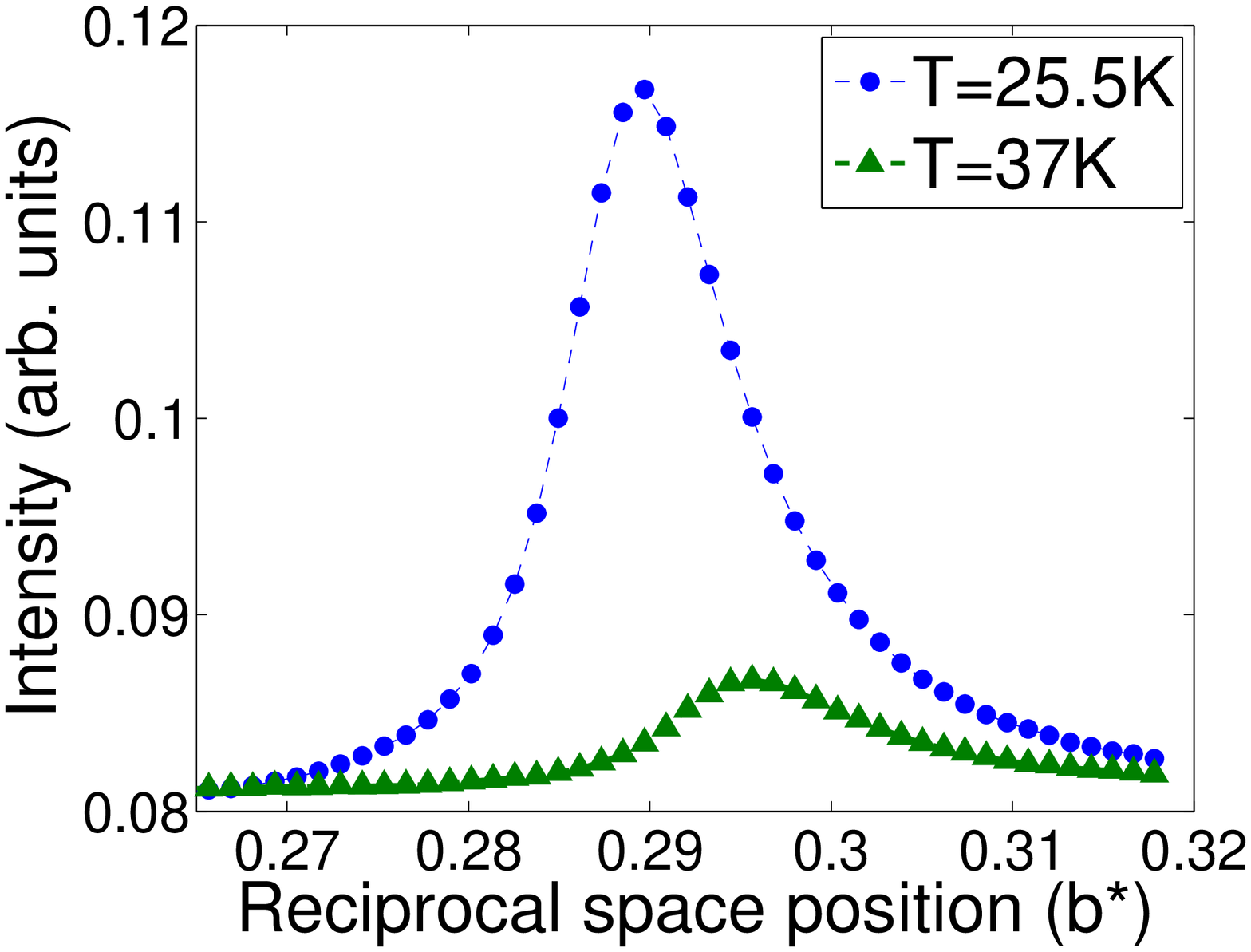}}
    \hspace{-0.3cm}
\subfigure[]{\label{L$_2$ peak}
    \includegraphics[width=.25\textwidth,bb=10 170 570 600]{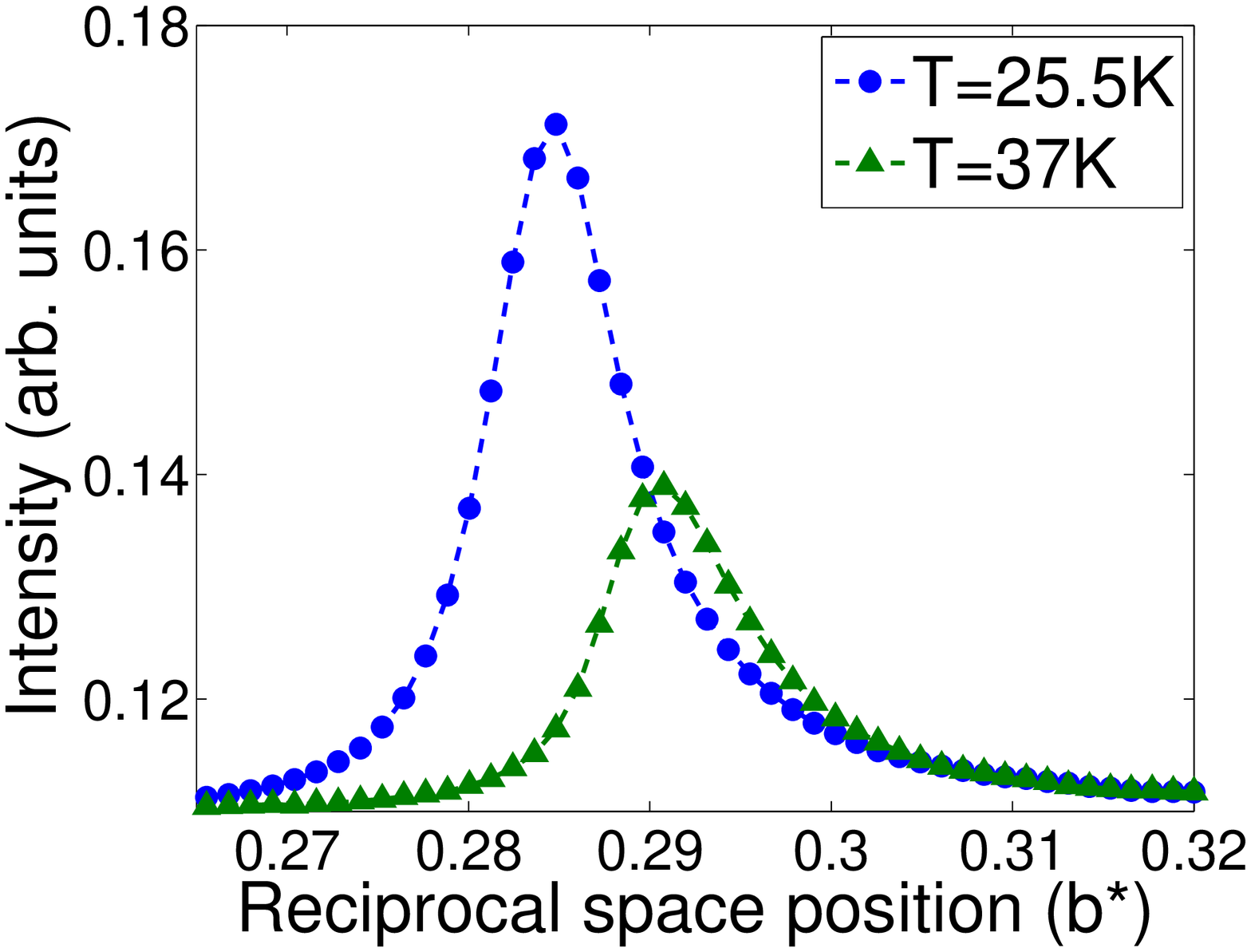}}\\
    \vspace{-0.35cm}
\subfigure[]{\label{Escan}
\includegraphics[width=.5\textwidth,bb=5 180 595 600,clip]{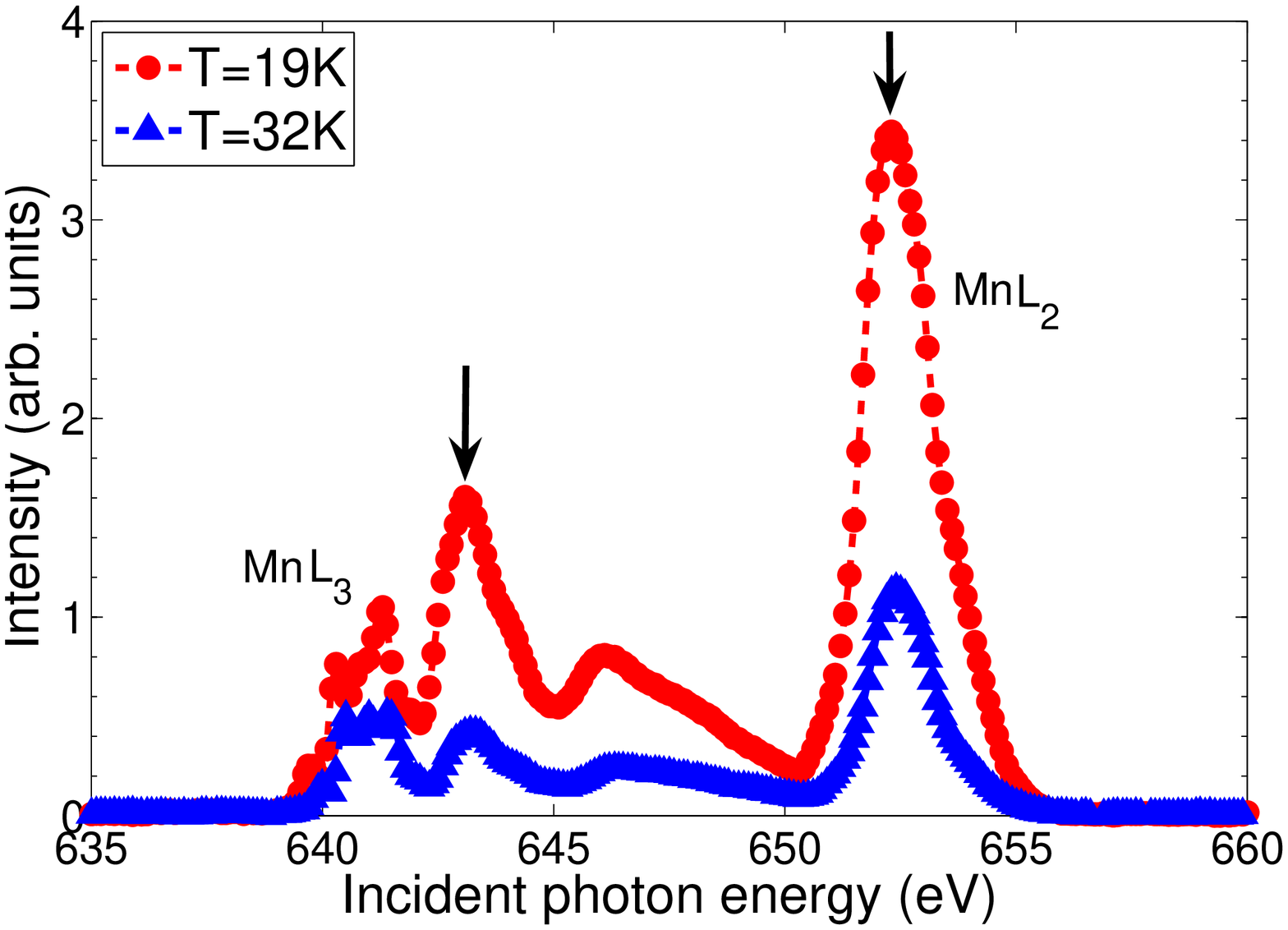}}
\caption{\label{fig1}$\theta$-2$\theta$ scan of the (0 $q$ 0) reflection
in the cycloidal (25.5K) and collinear (32K) phases collected with
incident X-ray photon energies equal to (a) the Mn $L_3$ resonance
(639 eV) and (b) the Mn $L_2$ resonance (652 eV) as determined from (c)
an energy scan at fixed wavevector of the (0 q 0) reflection in the
two phases. The vertical arrows indicate the energies at which a temperature dependence was recorded.}
\end{figure}

We first consider the results for the Mn $L$ edges taken using the
[0~1~0] orientated sample. On cooling below $T_{N1}\approx$ 42~K an 
$F$-type satellite diffraction peak was observed at (0~q~0), with
q $\approx$ 0.295~b* just below  $T_{N1}$. The peak was present at
both the $L_2$ and $L_3$ edges, and was found to increase in
intensity and move to lower q as the temperature was decreased
(Fig.\ \ref{L$_3$ peak} and \ref{L$_2$ peak}). By fitting this satellite peak to a Lorentzian
line shape, the correlation lengths (defined as $\zeta=\frac{1}{\kappa}$, where $\kappa$ is the characteristic half width of the Lorentzian distribution in reciprocal lattice units) were determined to exceed 200{\AA} at both the $L_2$ and $L_3$ edges, and in both the cycloidal and collinear phases. This result demonstrates that the X-rays probe a significant number of unit cells within the crystal and hence, these measurements are not particularly surface sensitive.

A scan of the incident photon energy at fixed wavevector transfer in the
high-temperature collinear phase revealed strong enhancements of the
scattering cross-section at the Mn $L_2$ and $L_3$ edges (Fig.\
\ref{Escan}). While a simple, single resonant response is evident at
the $L_2$ edge, the energy line shape displays much more structure in
the vicinity of the $L_3$ edge. 
Notwithstanding these important
details, the strong electric dipole resonances, combined with the sharpness
of the peaks in reciprocal space, establishes the fact that the Mn
$3d$ electronic states display long-range order in the collinear phase. In
Fig.\ \ref{Escan} the results of an energy scan in the cycloidal
phase at 19~K are also shown. Apart from an overall increase in
intensity, the response in this phase is indistinguishable from that
in the collinear one. 

It should be noted that a number of soft X-ray resonant scattering studies have been made of related rare-earth manganite compounds, which are not multiferroic \cite{Wilkins03_1,Wilkins03_2,wilkinsLSMO,Thomas}. Results from all of these studies have shown that, for both the magnetic and orbital reflections, the resonant feature at the Mn $L_3$ edge is always strongest. This is clearly not the case for the (0 $q$ 0) reflection observed in this study. However, to obtain further information from this fixed wavevector energy scan, detailed modeling of the electronic structure is required, which is beyond the scope of the present work.

\begin{figure}[t]
\centering
\includegraphics[width=.6\textwidth,bb=20 165 510 680,clip]{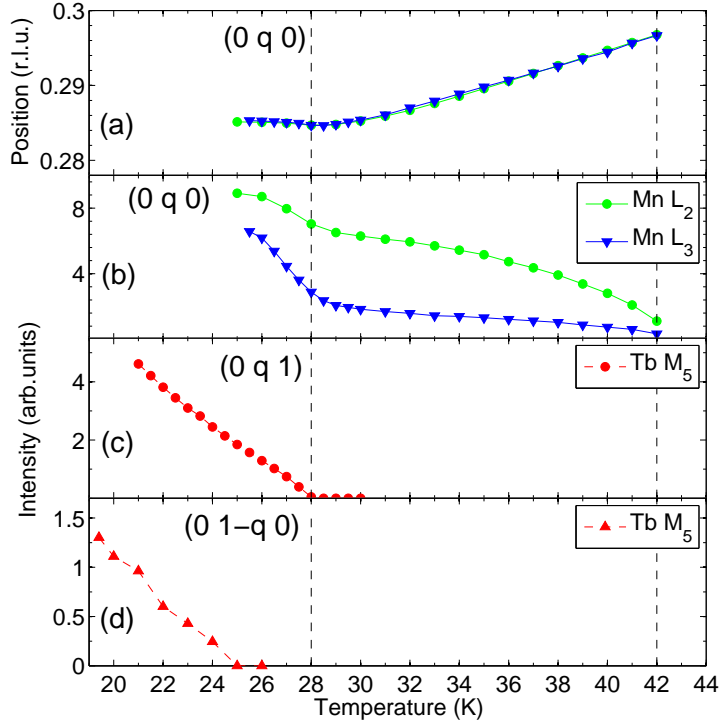}
\caption{\label{fig2}Temperature dependence of the (0 q 0)
superlattice reflection: (a) the position and (b) the integrated
intensity at the Mn $L_2$ (652.3 eV) and $L_3$ (643.1 eV) edges. (c)
the integrated intensity of the (0 q 1) superlattice reflection at
the Tb $M_5$ edge (1240 eV), (an offset of -2K has been applied to
this data). Finally (d) is the integrated intensity of the (0 1-q 0)
superlattice reflection, again this was recorded with photons equal
in energy to the Tb $M_5$ edge (1240 eV).}
\end{figure}

Scans parallel to {\bf b*}, across the (0 q 0) reflection were performed as
a function of temperature at incident photon energies equal to the
main features of the Mn $L_2$ and $L_3$ resonances shown in Fig.\
\ref{Escan}. Figure~\ref{fig2} shows the temperature dependence of
(a) the position and (b) the integrated intensity of the peak
obtained by fitting a Lorentzian function to the peak
profiles. In the cycloidal/ferroelectric phase below 28~K, the
propagation vector is only weakly temperature dependent with
$q\simeq0.285$ b* before increasing linearly with increasing
temperature within the collinear phase to a maximum value of
$q\simeq0.295$ b* at $T=42$ K. This trend is consistent with that
deduced by neutron and hard X-ray scattering experiments
\cite{kenzelmann,mannix}. Above 42 K the peak was not observed. The
evolution of the integrated intensity as a function of temperature
clearly differs between the measurements performed at the Mn $L_2$
and $L_3$ edges, with a more significant change at the $L_3$ edge at
the transition between the two magnetic phases. 
In other words the $L_2$/$L_3$ branching ratio is temperature dependent.
A strongly temperature dependent branching ratio has previously been observed in DyFe$_4$Al$_8$, 
which was attributed to the effect of magneto-elastic coupling \cite{Langridge}.
 
The fact that the main resonances occur at the Mn $L_2$ and $L_3$
edges indicates that the electric dipolar (E1) resonance dominates and therefore the Mn $3d$ electronic states are being probed.
When this is combined with the observation, that the thermal evolution of
the wavevector shown in Figure~\ref{fig2}(a) tracks that of the
fundamental magnetic wavevector {\bf q} determined by neutron
diffraction, the conclusion that the XRS in this experiment is most probably magnetic in origin may be drawn.
The only other possible order parameter that might be
probed using XRS for a purely dipolar transition is that associated
with ordering of the electric quadrupole (orbital ordering) which these experiments do not exclude.

\begin{figure}
\centering \subfigure[]{\label{M$_5$ peak}
    \includegraphics[width=.25\textwidth,bb=20 180 555 610]{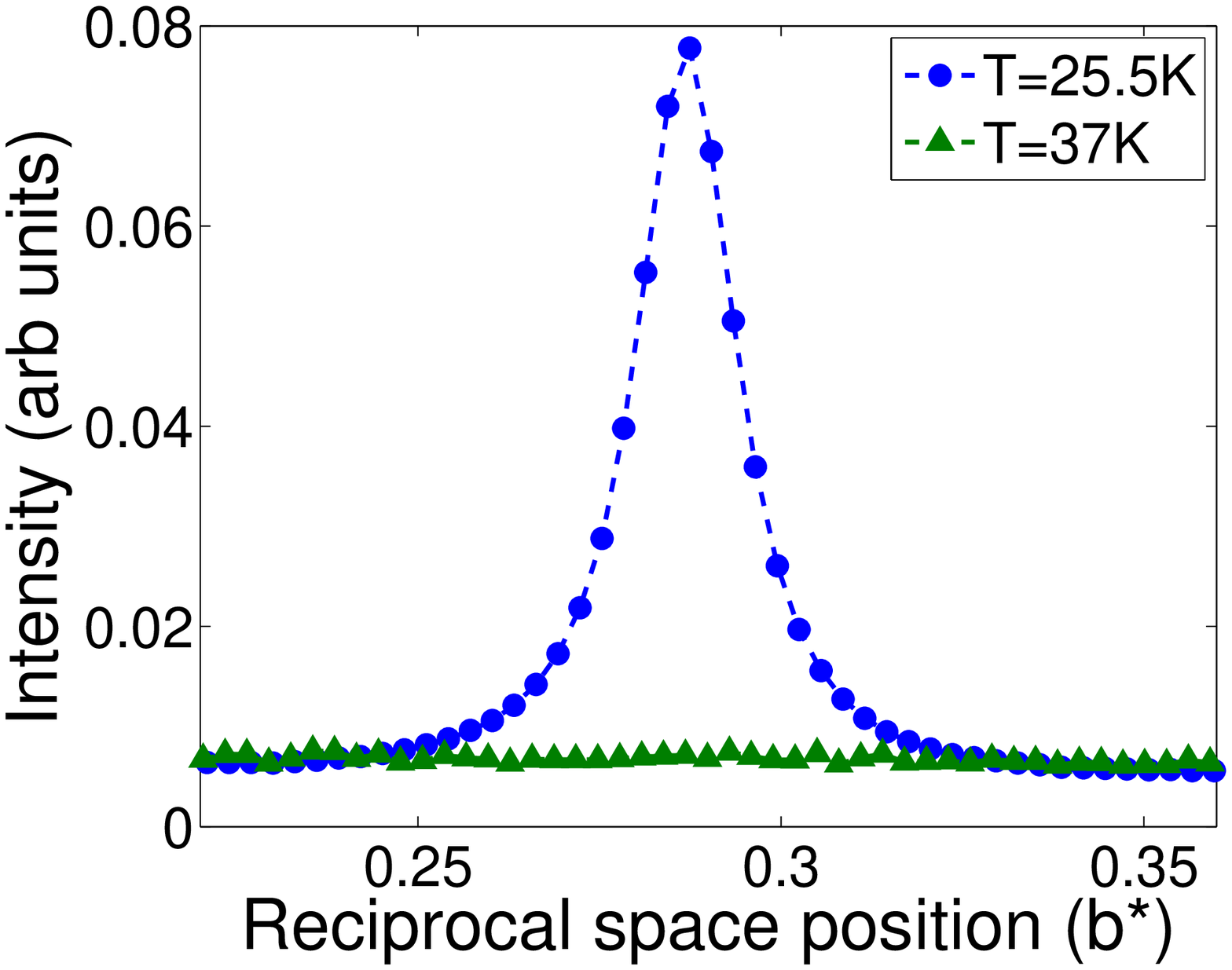}}
    \hspace{-0.3cm}
\subfigure[]{\label{M$_4$ peak}
    \includegraphics[width=.25\textwidth,bb=20 180 555 610]{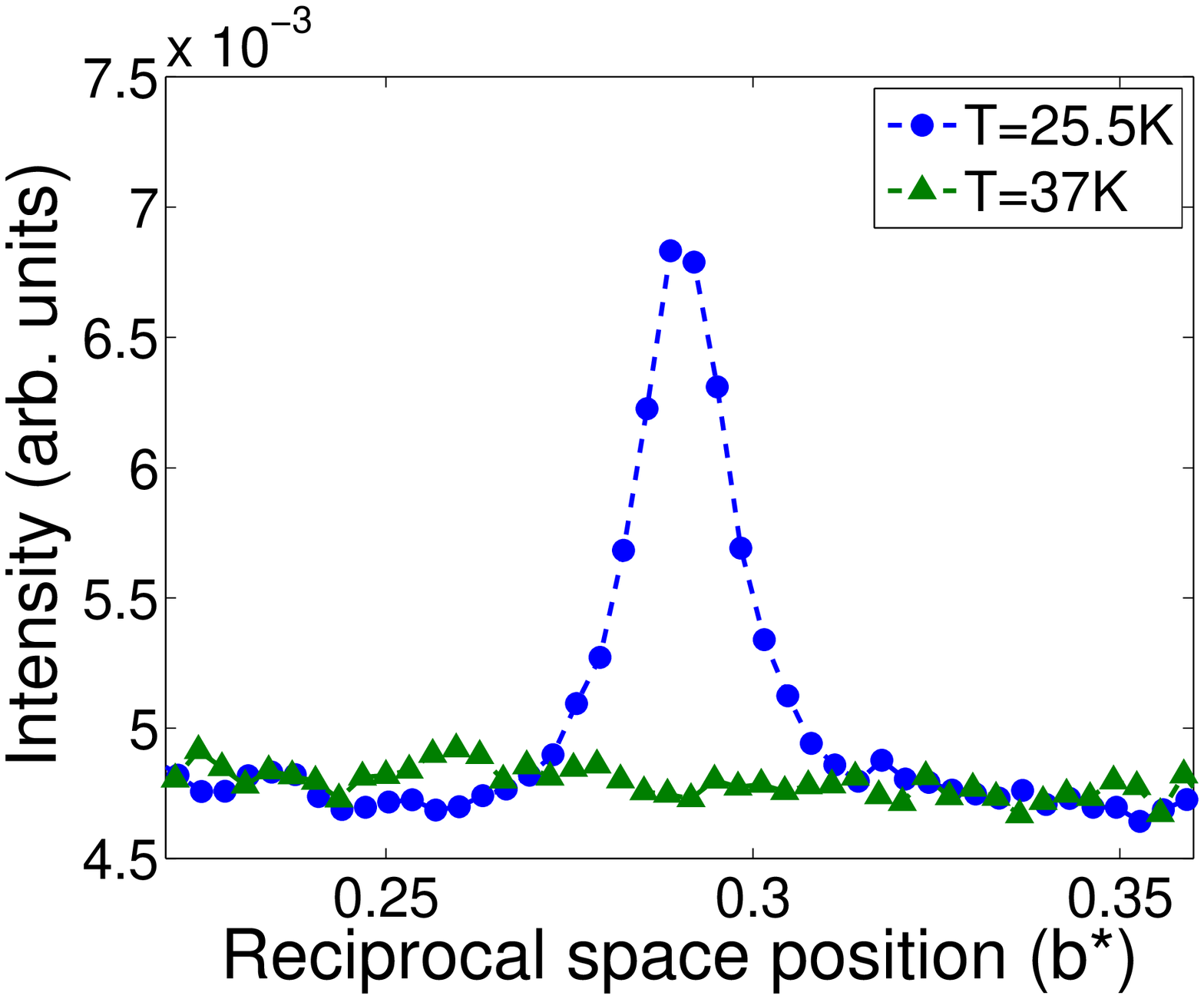}}\\
    \vspace{-0.35cm}
\subfigure[]{\label{Escan2}
\includegraphics[width=.5\textwidth,bb=0 170 565 610,clip]{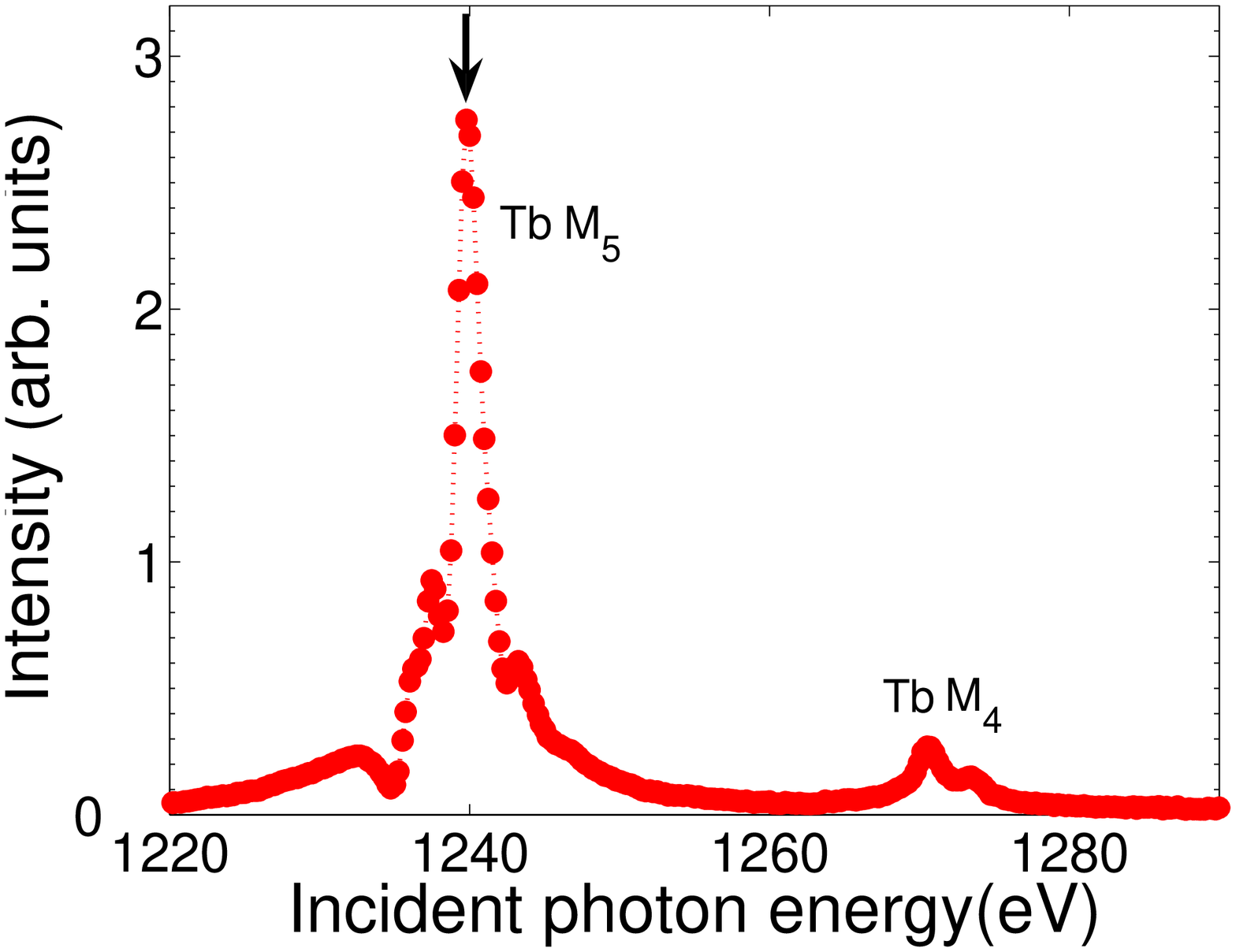}}
\caption{\label{fig3}$\theta$-2$\theta$ scan of the (0 q 1) reflection in
the cycloidal phase at $T=26$ K collected with incident X-ray photon
energies equal to (a) the Tb $M_5$ resonance (1240 eV) and (b) the
Tb $M_4$ resonance (1271 eV) as determined from (c) an energy scan
at fixed wavevector of the (0 q 1) reflection. The vertical arrow indicate the energy at which a temperature dependence was recorded.}
\end{figure}

We now consider the results taken with photon energies close to the Tb
$M_4$ \& $M_5$ edges. In contrast to the Mn $L$ edges, where only the F-type reflection may be observed, at the Tb $M$ edges, A and C type reflections are also accessible. (To observe the A-type reflection, the [0~1~0] was replaced by the [0~0.28~1] orientated sample.) Comprehensive searches in the collinear phase for any of these reflections produced negative results. On cooling into the cycloidal phase, strong, well-defined A-type reflections appeared at both the $M_4$ and $M_5$ edges. The sharpness of the diffraction profiles (Fig. \ref{M$_4$ peak} and \ref{M$_5$ peak}) immediately establishes that the electronic states (in this case the Tb 4f states) are long-range ordered.
Figure~\ref{Escan2} shows a energy scan at the fixed wavevector of the (0~q~1) reflection in the cycloidal phase, 
demonstrating significant resonances at the $M_5$ edge and $M_4$ edges. This indicates that for
the (0 q 1) reflection, the Tb $4f$ electronic states are strongly
influenced by the cycloidal magnetic order.
Figure~\ref{fig2} (c) shows the temperature dependence of the
scattered intensity of the $A$-type (0 q 1) reflection taken at the
Tb $M_5$ edge. As for the $F$-type peak, where the results were taken at the Mn $L$ edges, the magnitude of the
modulation wavevector evolves as a function of temperature.
For this crystal orientation, however, the reflection becomes off-specular.
Experimental limitations at 5U1 made it impossible to accurately resolve the evolution of 
peak's position as the temperature was increased, and hence here we only present the
integrated intensity as a function of temperature, determined as for
the (0 q 0) reflection. As can be seen, the intensity of the
scattering at the $M_5$ edge drops rapidly and linearly with
increasing temperature, with zero intensity being observed for
$T\geq30$ K. Hence, it would appear that in this domain the
localized Tb $4f$ states are strongly affected by the cycloidal
order, but not by the collinear magnetism.

For the $F$-type (0 q 0) reflection, a very weak resonance was
observed in the cycloidal phase at the Tb $M_5$-edge, but only with
$\pi$-polarized incident photons. However, a stronger Tb $M_5$ resonance
reflection was observed at a position of (0 $\sim$0.78 b* 0). This
superlattice reflection corresponds to the $C$-type (0 1-q 0)
domain. Like the $F$-type reflection, the peak was only observed
with $\pi$-polarized incident photons. Figure ~\ref{fig2} (d) shows the temperature
dependence of the scattered intensity for this (0 1-q 0) peak. 
As the temperature was increased, the intensity of the scattering at
the Tb $M_5$ edge decreased, with zero intensity being observed for $T\geq24$ K.
The position of this peak remained constant as the temperature was increased. 

Finally, we note that fixed wavevector energy scans for the
$F$-type (0 $q$ 0) reflection failed to identify any clear response
in the vicinity of the oxygen $K$ (543.1 eV: $1s\rightarrow 2p$) edge.

\section{Conclusions}
In conclusion, we have performed the first direct element-specific
study of the effect of the magnetic order on the electronic
structure of magnetoelectric multiferroic TbMnO$_3$. We have demonstrated that the Mn $3d$ localized bands display strong long range order in both magnetic phases for the F-type domain. The temperature dependence of this (0 q 0)
reflection is in good agreement with previously observed trends for
both the position of the modulation wavevector (0 $q_{Mn}$ 0) and the scattered intensity,
with clear changes at the magnetic phase transitions.
Mn $L$-edge energy scans at this wavevector show minimal changes in the overall $3d$ band 
structure between the collinear and cycloidal phases. The scattered intensity as a function of temperature does differ, however, between measurements performed at the Mn $L_2$ and $L_3$ edges. The transition at 28K into the ferroelectric phase is more significant for the measurements performed at the Mn $L_3$ edge.
The energy scans taken at the Tb $M$-edges clearly show, that for the A-type (0 q 1)
reflection, the Tb $4f$ band is highly influenced by the cycloidal magnetic order, whilst this reflection
was absent in the collinear phase, indicating that there is no long range ordering of the Tb $4f$ states for this phase.
This data supports the neutron diffraction model which states that the Tb $4f$ states should be disordered in the collinear phase. 
However the absence of Tb $4f$ ordering in the collinear phase suggests that ordering of Tb $5d$ bands as seen with hard X-rays \cite{mannix} is of a different origin.
In addition to strong Tb $M$ edge resonances observed for the A-type peak, much weaker Tb $M$ edge resonance peaks corresponding to the F-type (0 q 0) and C-type (0 1-q 0) reflections were observed in the cycloidal phase. The fact that these two reflections were observed with $\pi$-incident X-rays
only and are much weaker, suggest a difference between the magnetic structures of the domain states.
Finally, the lack of an F-type reflection in the vicinity of the oxygen $K$ edge, shows that for this reflection at least, there is no long range ordering of the oxygen $2p$ band.

\ack
The authors thank R. Bean for his experimental assistance. 
Work in London was supported by the EPSRC and a Wolfson Royal Society Award and in Durham and Oxford by the EPSRC.
The work at Brookhaven National Laboratory is supported by the Office of Science, U.S. Department of Energy, under contract no. DE-AC02-98CH10886.
\appendix

\section*{References}
\bibliographystyle{unsrt}
\bibliography{tbmno3_arXiv}

\begin{thebibliography}{10}

\bibitem{eerenstein}
W.~Eerenstein, N.~D. Mathur, and J.~F. Scott.
\newblock {\em Nature}, 442:759, 2006.

\bibitem{Fiebig}
M.~Fiebig.
\newblock {\em J. Phys. D: Applied Physics}, 38:R123--R152, 2005.

\bibitem{Spaldin}
N.~A. Spaldin and M.~Fiebig.
\newblock {\em Science}, 309:391--392, 2005.

\bibitem{Hill}
N.A. Hill.
\newblock {\em J. Phys. Chem. B}, 104:6694, 2000.

\bibitem{Cheong}
S.-W. Cheong and M.~Mostovoy.
\newblock {\em Nature Materials}, 6:13, 2007.

\bibitem{KimuraTMO}
T.~Kimura, T.~Goto, H.~Shintani, K.~Ishizaka, T.~Arima, and Y.~Tokura.
\newblock {\em Nature}, 426:55, 2003.

\bibitem{yamasaki}
Y.~Yamasaki, H.~Sagayama, T.~Goto, M.~Matsuura, K.~Hirota, T.~Arima, and
  Y.~Tokura.
\newblock {\em \PRL}, 98:147204, 2007.

\bibitem{kenzelmann}
M.~Kenzelmann, A.~B. Harris, S.~Jonas, C.~Broholm, J.~Schefer, S.~B. Kim, C.~L.
  Zhang, S.-W. Cheong, O.~P. Vajk, and J.~W. Lynn.
\newblock {\em \PRL}, 95:087206, 2005.

\bibitem{ewings}
R.~A. Ewings, A.~T. Boothroyd, D.~F. McMorrow, D.~Mannix, H.~C. Walker, and
  B.~M.~R. Wanklyn.
\newblock {\em \PR B}, 77:104415, 2008.

\bibitem{Prokhnenko}
O.~Prokhnenko, R.~Feyerherm, E.~Dudzik, S.~Landsgesell, N.~Aliouane, L.~C.
  Chapon, and D.~N. Argyriou.
\newblock {\em \PRL}, 98:057206, 2007.

\bibitem{yang}
C.-H. Yang, J.~Koo, C.~Song, T.~Y. Koo, K.-B. Lee, and Y.~H. Jeong.
\newblock {\em \PR B}, 73:224112, 2006.

\bibitem{koo}
J.~Koo, C.~Song, S.~Ji, J.-S. Lee, J.~Park, T.-H. Jang, C.-H. Yang, J.-H. Park,
  Y.~H. Jeong, K.-B. Lee, T.~Y. Koo, Y.~J. Park, J.-Y. Kim, D.~Wermeille, A.~I.
  Goldman, G.~Srajer, S.~Park, and S.-W. Cheong.
\newblock {\em \PRL}, 99(19):197601, 2007.

\bibitem{matteo}
S.~Di Matteo, Y.~Joly, and C.~R. Natoli.
\newblock {\em \PR B}, 72:144406, 2005.

\bibitem{VanAken}
B.~Van Aken, J-P. Rivera, H.~Schmid, and M.~Fiebig.
\newblock {\em Nature}, 449:702, 2008.

\bibitem{mannix}
D.~Mannix, D.~F. McMorrow, R.~A. Ewings, A.~T. Boothroyd, D.~Prabhakaran,
  Y.~Joly, B.~Janousova, C.~Mazzoli, L.~Paolasini, and S.~B. Wilkins.
\newblock {\em \PR B}, 76:184420, 2007.

\bibitem{argyriou}
D.~N. Argyriou, N.~Aliouane, J.~Strempfer, I.~Zegkinoglou, B.~Bohnenbuck,
  K.~Habicht, and M.~v.~Zimmermann.
\newblock {\em \PR B}, 75:020101, 2007.

\bibitem{Quezel}
S.~Quezel, F.~Tcheou, J.~Rossatmignod, G.~Quezel, and Roudaut E.
\newblock {\em Physica B \& C}, 86:916, 1977.

\bibitem{Blasco}
J.~Blasco, C.~Ritter, J.~Garc\'ia, J.~M. de~Teresa, J.~P\'erez-Cacho, and M.~R.
  Ibarra.
\newblock {\em \PR B}, 62:5609--5618, 2000.

\bibitem{Kajimoto}
R.~Kajimoto, H.~Yoshizawa, H.~Shintani, T.~Kimura, and Y.~Tokura.
\newblock {\em \PR B}, 70:012401, 2004.

\bibitem{Wilkins03_1}
S.B. Wilkins, P.D. Spencer, P.D. Hatton, S.P. Collins, M.D. Roper,
  D.~Prabhakaran, and A.T. Boothroyd.
\newblock {\em \PRL}, {91}({16}):167205, {2003}.

\bibitem{Wilkins03_2}
S.B. Wilkins, P.D. Hatton, M.D. Roper, D.~Prabhakaran, and A.T. Boothroyd.
\newblock {\em \PRL}, {90}({18}):187201, {2003}.

\bibitem{wilkinsLSMO}
S.B. Wilkins, N.~Stojic, T.~A.~W. Beale, N.~Binggeli, C.~W.~M. Castleton,
  P.~Bencok, D.~Prabhakaran, A.~T. Boothroyd, P.~D. Hatton, and M.~Altarelli.
\newblock {\em \PR B}, 71:245102, 2005.

\bibitem{Thomas}
K.~J. Thomas, J.~P. Hill, S.~Grenier, Y-J. Kim, P.~Abbamonte, L.~Venema,
  A.~Rusydi, Y.~Tomioka, Y.~Tokura, D.~F. McMorrow, G.~Sawatzky, and M.~van
  Veenendaal.
\newblock {\em \PRL}, 92(23):237204, 2004.

\bibitem{Langridge}
S.~Langridge, J.A. Paixao, N.~Bernhoeft, C.~Vettier, G.H. Lander, D.~Gibbs,
  S.A. Sorensen, A.~Stunault, D.~Wermeille, and E.~Talik.
\newblock {\em \PRL}, {82}({10}):{2187--2190}, {1999}.

\end{thebibliography}

\end{document}